# Controlling Photon Transverse Orbital Angular Momentum in High Harmonic Generation

Yiqi Fang[1], Shengyue Lu[1,2] and Yunquan Liu[1,2,3,4]

[1]*State Key Laboratory for Mesoscopic Physics and Collaborative Innovation Center of Quantum Matter, School of Physics, Peking University, Beijing 100871, China*
[2]*Beijing Academy of Quantum Information Sciences, Beijing 100193, China*
[3]*Collaborative Innovation Center of Extreme Optics, Shanxi University, Taiyuan, Shanxi 030006, China*
[4]*Peking University Yangtze Delta Institute of Optoelectronics, Nantong, Jiangsu, 226010, China*

**Abstract:** High harmonic generation (HHG) with longitudinal optical orbital angular momentum has attracted much attention over the past decade. Here, we present the first study on the HHG with transverse orbital angular momentum driven by the spatiotemporal optical vortex (STOV) pulses. We show that the produced spatial-resolved harmonic spectra reveal unique structures, such as the spatially spectral tilt and the fine interference patterns. We show these spatio-spectral structures originate from both the macroscopic and microscopic effect of spatiotemporal optical singularity in HHG. Employing two-color counter-spin and counter-vorticity STOV pulses, we further discuss a robust method to control the spatiotemporal topological charge and spectral structure of high-order harmonics. The conservation rule of photon transverse orbital angular momentum in HHG process is also discussed when mixing with photon spin angular momenta.

Optical vortices are the localized fields where the energy flow of electromagnetic field circulates around a local axis, leading to a null in the field amplitude and a singularity in the field phase. The optical vortex endues the light fields with the well-known optical orbital angular momentum (OAM) of $\ell\hbar$ per photon ($\ell$ is the topological charge) [1]. The phase winding of a conventional optical vortex resides in the spatial domain, forming a longitudinal OAM parallel to the propagation direction of laser. In the past three decades, the longitudinal OAM of photons has been widely studied and employed in various fields [2-5]. Recently, the spatiotemporal optical vortex (STOV), an analog of optical vortex in space-time plane, has attracted increasing attention in optics [6-11]. The transverse OAM perpendicular to the propagation direction of STOV provides a new degree of freedom for optical modulations and further applications. The STOV was first experimentally observed from the filamentation processes in material media [6] and was then successfully realized in free space by modulating the laser phase in spatial ($x$) and spectral ($\omega$) domain [7,8]. Very recently, the conversion of STOV is revealed and characterized in second harmonic generation process [12,13].

High harmonic generation (HHG) is an extreme optical non-linear process, which occurs when intense ultrashort laser pulses interact with matters. Microscopically, HHG is typically dictated by an ionization-and-recollision picture [14], accompanied by the emission of up-converted high-energy photons [15,16]. Macroscopically, the coherence among the photons radiated from different target-atoms is critical, which plays an important role in controlling the extreme ultraviolet (EUV) radiations. Today, the attosecond pulses with HHG have been widely applied in probing ultrafast dynamics [17-20] and imaging nanoscale structures [21,22]. In the past decade, the emerging OAM-driven HHG has gained much exciting progress for shaping the EUV radiations [23-31]. These studies provide a powerful toolbox for extensive applications, such as controlling the spin-orbit state of attosecond pulses [26] and generating EUV beams with time-varying OAM [27]. However, since the electron ponderomotive motion is much smaller than the optical wavelength of driving laser, the spatial phase structure of conventional vortex beams has less effect on the microscopic process of HHG [32]. The generation of EUV vortices in these previous studies is simply attributed to the macroscopic effect of HHG [24,30,31]. Because of this flaw, the manipulation of the harmonic radiations by using optical OAM is fundamentally limited. So far, all of the studies have been concentrated on HHG with the longitudinal OAM. Compared with the conventional optical vortex, the newly discovered STOV possesses the vortex structure in space-time plane, making it possible to imprint the optical OAM on both microscopic and macroscopic effects of HHG. HHG with the transverse OAM would have a promising perspective for both attosecond science and extreme optics.

In this Letter, we study the conversion and modulation of photon transverse OAM in HHG. We show that, (i) the spatial chirp of the fundamental STOV leads to a significant tilt for the harmonics in spatial-resolved HHG spectrum, (ii) the spatiotemporal singularity has a notable effect on the electron sub-cycle dynamics, giving rise to the unusual fine interference structures in the spectrum. This has been never shown up in the previous HHG phenomenon. (iii) the transverse OAM per photon $L_n$ of the $n$th harmonic is equal to $n$ times that of the fundamental STOV, i.e., $L_n = n\ell\hbar$, (iv) by employing two-color counter-spin and counter-vorticity STOV pulses, the spectral feature and spatiotemporal topological charge (transverse OAM) of the EUV STOVs can be efficiently controlled and selected. The conservation rule of photon transverse OAM in HHG process is also discussed when mixing with photon spin angular momenta.

The fundamental STOV pulses can be described as [6,7],

$$E(x,y,z,t) = E_0 A_0 \cdot \left[ \left(\xi/w_\xi\right)^2 + \left(x/w_x\right)^2 \right]^{|\ell|/2} \exp\left[-(x^2/w_x^2 + y^2/w_y^2 + \xi^2/w_\xi^2)\right] \\ \times \exp\left[i(-\ell\varphi_{st} + \omega_0 t + k_0 z)\right] \quad (1)$$

where $(x, y, z, t)$ are the spatial and time coordinates, $\xi = ct - z$ is a longitudinal coordinate local to the pulse, $c$ is the group velocity of light, $E_0$ is the peak electric field strength, $A_0$ is a normalization constant, $w_\xi$ and $w_x$ are the spatial scale widths, $\varphi_{st} = \tan^{-1}(xw_\xi/\xi w_x)$ is the azimuthal angle in space-time plane (or, equivalently, the $x$-$z$ plane), and $\omega_0$ and $k_0$ are the center angular frequency and wavenumber, respectively. In Fig. 1(a), we show the spatiotemporal electric field structure of the fundamental STOV of $\ell = 1$ at $y = 0$. In the calculation, the fundamental field of 400-nm wavelength ($\lambda_{400}$) with the peak intensity of $5\times10^{14}$ W/cm$^2$ is employed, and $w_\xi$ and $w_x$ are both taken to be $4\lambda_{400}$. As shown, the electric field distribution reveals a fork grating structure [33], and the electric field oscillates with diverse frequencies at different $x$-positions. By performing the Fourier transform of time, we can obtain the angular-frequency spectrum of the laser field at each $x$-position. For $\ell = 1$, the Fourier transform result is analytically given by

$$F(\omega, x) = \frac{\sqrt{\pi} E_0 A_0 w_\xi}{c} \exp[-(x/w_x)^2]$$
$$\times \exp[-\frac{1}{4}(w_\xi/c)^2(\omega-\omega_0)^2]\left[-\frac{1}{2}\frac{w_\xi}{c}(\omega-\omega_0) - \frac{x}{w_x}\right]. \qquad (2)$$

The spatio-spectral weight of each angular frequency is then given by $|F(x, \omega)|$.

In Fig. 1(b), we present the distribution of $|F(x, \omega)|$ in a pseudo-color map. The calculated result shows the polychromatic nature of STOV pulses with transverse OAM. If comparing the distribution of $|F(x, \omega)|$ with the center angular frequency of the STOV pulses (i.e., $\omega_0 = 0.114$ a.u.) [the white solid line in Fig. 1(b)], the field spatial chirp can be observed. To quantify such spatial chirp, we define a local center angular frequency, given by

$$\omega(x) = (\int_{-\infty}^{\infty} |F(x,\omega)| \omega d\omega) / (\int_{-\infty}^{\infty} |F(x,\omega)| d\omega) = \omega_0 + \Delta\omega(x). \qquad (3)$$

The physical meaning of Eq. (3) is very clear, in which $\Delta\omega(x) = (cw_x/w_\xi x)[1 - 1/(1 + \frac{\sqrt{\pi}x^2}{w_x^2} e^{x^2/w_x^2})]$ quantifies the local spatial chirp of the fundamental STOV. As shown in Fig. 1(b) (the black dashed curve), the local field is blue-shifted for $x > 0$ and the local field is red-shifted for $x < 0$. Interestingly, we find such spatial chirp is somewhat similar to that in the attosecond lighthouse technique [34,35], which has been realized to produce angularly separated attosecond pulses by rotating the wavefront of driving field.

Then, we employ the integral-form SFA method [36] to simulate the HHG process with the aforementioned fundamental STOV pulses. Here, we focus on the harmonic spectrum near the cut-off region. We show the simulated spatial-resolved harmonics spectrum in Fig. 1(c). At the first glance, the harmonic emission is spatially tilted in the spectrum. The local center angular frequency of the $n$th harmonic can be related to that of the fundamental STOV pulses with

$$\omega_n(x) = n\omega(x) = n\omega_0 + n\Delta\omega(x). \qquad (4)$$

According to Eq. (4), the spatio-spectral tilt of harmonics is described by $n\Delta\omega(x)$, which scales linearly with the harmonic order. We plot $\omega_n(x)$ in Fig. 1(c) (the black dashed curves). Remarkably, it has a nice correspondence with the spatio-spectral tilt of harmonics calculated with the SFA. This means that the spatio-spectral tilt of harmonics results from the spatial chirp of the fundamental STOV pulses.

Moreover, one can find that there are fine interference structures in each harmonic, and the number of interference fringes is associated with the harmonic order. The tilted interference fringes imply that the frequency and *x*-position are inherently entangled in the formation of such interference structures in HHG with spatiotemporal singularity. To our knowledge, this is a unique spatio-spectral interference structure of HHG driven by the STOV.

To reveal the origin of the fine interference structures, we then analyze the temporal character of HHG bursts. Since the phase singularity induces a dark core centered at *t* = 0 and *x* = 0 in the light field, the HHG bursts are naturally divided into two clusters in time domain, as shown in Fig. 2(a). We categorize the harmonic bursts emitted before the phase singularity as cluster-A and after the phase singularity as cluster-B. In theory, we resort to the Gabor transform to calculate the HHG spectrum from these two groups of HHG bursts, individually. The calculated results for *x* = 0 and *x* = $\lambda_{400}$ are presented in Figs. 2(b) and 2(c), respectively. One can notice that there is no interference pattern in the spectrum from the individual cluster-A or cluster-B. Only when these two clusters are both considered, the fine interference patterns show up. This indicates that the fine interference patterns come from the interference between the harmonic bursts emitted before the phase singularity and the bursts emitted after the phase singularity. Such a mechanism can be analogous to a temporal double-slit [37].

To further identify the underlying mechanism of the conversion of transverse OAM in HHG, we have developed an analytical model based on the quantum-orbit model of the SFA framework [38]. In the quantum-orbit model, the HHG spectrum *P*(*x*, *ω*) is given by the sum over different quantum orbits *D*(*x*, *ω*)

$$P(x,\omega) \propto \omega^4 \left| \sum_m D(x,\omega) \right|^2. \tag{5}$$

Here, we derived a simple and analytical expression for the amplitude of each quantum orbit :

$$D(x,\omega) \propto (-1)^m \left| \frac{(2I_p)^2}{E(x,t_m^{(i)})} \right|^{2/\sqrt{2I_p}-1} \frac{\sqrt{(\omega-I_p)}}{I_p} e^{-\frac{2(2I_p)^{3/2}}{3E(x,t_m^{(i)})} - \frac{4\omega}{5I_p}} e^{i\omega t_m^{(i)}}, \tag{6}$$

in which $t_m^{(i)} = (0.297 + m\pi + \varphi_{st})/\omega_0$ is the ionization moment of *m*th orbit, and $I_p$ is the ionization potential of target atoms. The detailed derivation has been presented in [39]. Compared with the original quantum-orbit model, this formula is very concise, and it allows one to calculate the HHG spectrum without solving the fussy saddle point equation [43]. We employ this model to simulate the HHG process. The simulated result (Fig. 2(d)) has a good agreement with the result from the integral-form SFA (Fig. 1(c)).

As known, the interference patterns are determined by the phase of waves. In this analytical model, the only phase factor of each orbit is contributed by the product between the harmonic angular frequency and the electron ionization moment. Since the electron recollision moment is directly related to the ionization moment, the STOV has significant effects on the ionization and recollision steps of the well-known three-step model [14]. When the STOV interacts with atoms in HHG, its spatiotemporal structure is first recorded by the ionization moment of electrons on sub-cycle time scale. The electron ionization moments are divided into two clusters and the temporal gap between these two clusters is $x$-dependent. After the electrons experience an excursion time in the external field, the special spatiotemporal distribution of electron ionization moments is remembered by the recombination moments of electrons. At the moment of recombination, such spatiotemporal structure is finally recorded by the emitted harmonic bursts, forming the unique HHG spectrum. Here, we would like to indicate that the influence of the fundamental STOV on the second step of three-step model, i.e., electron propagation, is trivial. Hence, though such influence is not considered in the analytical model, the calculated spectrum has a good agreement with the result from the SFA model.

Hitherto, it has been demonstrated that the conservation rule of the photon's energy, linear momentum [44], spin angular momentum [45], longitudinal OAM [24,25], and torus-knot angular momentum [30] are of utmost importance in stereo controlling the high harmonic radiations. The further intriguing question would be what is the effect of the transverse OAM in HHG. To answer this, we retrieve the spatiotemporal intensity and phase from the calculated harmonic fields. As shown in Fig. 3(a) and 3(b), the intensity distribution of the 17th harmonic reveals a typical doughnut-like structure and meanwhile, its phase distribution manifests as a phase winding of 17×2π. The spatiotemporal profiles of other order harmonics have been presented in [39]. Note that these structures exist in the space-time plane. The calculated results well indicate that the $n$th harmonic carries a spatiotemporal phase singularity whose topological charge is $n\ell$. To reveal the conservation rule of photon transverse OAM strictly, we further calculate the value of the transverse OAM per photon [39, 46]. The calculated OAM values per photon are $L_{13} = 12.9969\hbar$, $L_{15} = 14.9722\hbar$, $L_{17} = 16.9278\hbar$ and $L_{19} = 18.9010\hbar$ for the 13th, 15th, 17th, and 19th harmonics, respectively. The numerical results demonstrate the conservation of photon transverse OAM obeys the rule: $L_n =$

$n\ell\hbar$, where $L_n$ is the transverse OAM per photon of the $n$th harmonic. Besides, one can find that the harmonic intensity profile is spatially asymmetrical. Such unique feature of EUV STOV is caused by the spatial chirp of HHG spectrum [39].

Here, the large spatio-spectral tilt of harmonic spectrum will bring some difficulty to isolate different order EUV STOVs in practical applications. Besides, the STOV is unstable when the topological charge is large [8]. Thus, controlling the spatial spectrum and topology of EUV STOVs is an important issue. To realize this, we first attempt to employ two-color counter-vorticity STOVs ($\ell_{400}$ = 1 for 400-nm and $\ell_{800}$ = -1 for 800-nm) to drive the HHG. Since the spatiotemporal topological charges of the two-color STOVs are opposite, the spatial chirps of these two STOVs are inverse at each $x$-position. As a consequence, the spatio-spectral tilt of each harmonic is expected to be significantly amended, as shown in Fig. 4(a).

However, for the $n$th order harmonic, there are several allowed photon absorption channels ($n_{800}$, $n_{400}$), in which $n_{800}$ and $n_{400}$ are the numbers of photons absorbed from 800- and 400-nm fields. Here, the parity conservation requires that $n_{800} + n_{400}$ is odd and the energy conservation requires $n_{800} + 2n_{400} = n$. In principle, numerous solutions are satisfying these constraints. If taking into account the conservation rule of photon transverse OAM, the spatiotemporal topological charge of the $n$th harmonic is given by $\ell_n = \ell_{800} n_{800} + \ell_{400} n_{400}$. This implies that the transverse OAM mode is not pure for each harmonic. As shown in Fig. 4(a), the mixture of different transverse OAM modes will confuse the spatial-resolved harmonic spectrum.

To overcome this circumstance, we propose a two-color counter-spin and counter-vorticity STOVs scheme, in which the two-color fields are prepared with opposite spin angular momenta (SAM) and opposite transverse OAMs, i.e., $\sigma_{400}$ = 1, $\ell_{400}$ = 1 for 400-nm, and $\sigma_{800}$ = -1, $\ell_{800}$ = -1 for 800-nm ($\sigma$ is the spin state). With the additional spin constraints, the number of photons absorbed from each driving field must differ by one, so that the $3n$th harmonic is forbidden, the ($3n$+1)th harmonic is contributed by the channel ($n$+1, $n$), and the ($3n$-1)th harmonic is contributed by the channel ($n$-1, $n$) [47]. Now, there is only one channel allowed for each harmonic. Then, the spatiotemporal topological charges of ($3n$+1)th and ($3n$-1)th harmonics are given by ($n$+1)$\ell_{800}$ + $n\ell_{400}$ and ($n$-1)$\ell_{800}$ + $n\ell_{400}$, respectively. The spin state of ($3n$+1)th and ($3n$-1)th harmonics are given by ($n$+1)$\sigma_{800}$ + $n\sigma_{400}$ and ($n$-1)$\sigma_{800}$ + $n\sigma_{400}$. Both the topological charge and spin state are solely determined for each harmonic. We simulate the HHG with such two-color field. As shown in Fig. 4(b), the harmonics are now well separated. The calculated results verify that the topological charges of (3$n$+1)th and (3$n$-1)th harmonics equal -1 and +1, respectively. We also show the profiles of 17th harmonic in Figs. 4(c) and 4(d). The transverse OAM per photon of the 17th harmonic is exactly equal to 1.0$\hbar$.

Furthermore, we show that one can control the relative yield of (3$n$+1)th and (3$n$-1)th harmonics by tuning the intensity ratio ($I_{400}/I_{800}$) of the two-color counter-spin and counter-vorticity STOVs. In Figs. 5(a) and 5(b), we show the calculated HHG spectra for $I_{400}/I_{800}$ = 0.25 and 4, respectively. When the laser intensity of 800-nm is larger than that of 400-nm, the 10th harmonic is preferentially produced. Likewise, when the 400-nm field is stronger than 800-nm, the 11th harmonic is much more intense. This would allow one to control the overall spin state of attosecond pulse trains generated with HHG [48]. In our case, the spin state and spatiotemporal topology of the EUV STOVs are intrinsically coupled. Hence one can further manipulate the overall transverse OAM of the attosecond pulse trains. The calculated results demonstrate that, even in such complex laser field, the spectral shape of the HHG can be efficiently controlled.

In conclusion, we have studied the conversion and modulation of photon transverse OAM in HHG driven by STOV pulses. We have shown that the produced HHG radiations reveal many interesting features. These novel structures are generated by the joint contribution of the macroscopic and microscopic characters of spatiotemporal optical singularity in HHG. We have also demonstrated a robust scheme to control the topology and spectra of high harmonic STOVs. This work offers an intuitive physical picture of HHG process with transverse OAM, paving the way for the generation and manipulation of EUV STOV pulses. It is shown that the interplay between the spin state and the spatiotemporal orbit state of STOV is of utmost importance in HHG, which can be further employed in other nonlinear processes, such as second-harmonic generation [12,13]. The methodology can be also applied to modulate the photoelectron holography [49] and to reveal a new type of OAM-dependent dichroic photoelectric effect [50]. The conjunction of strong-field physics and STOV pulses, or even more complex structured light fields, will open a new and promising perspective in ultrafast science.

# Figure captions

# Figure 1

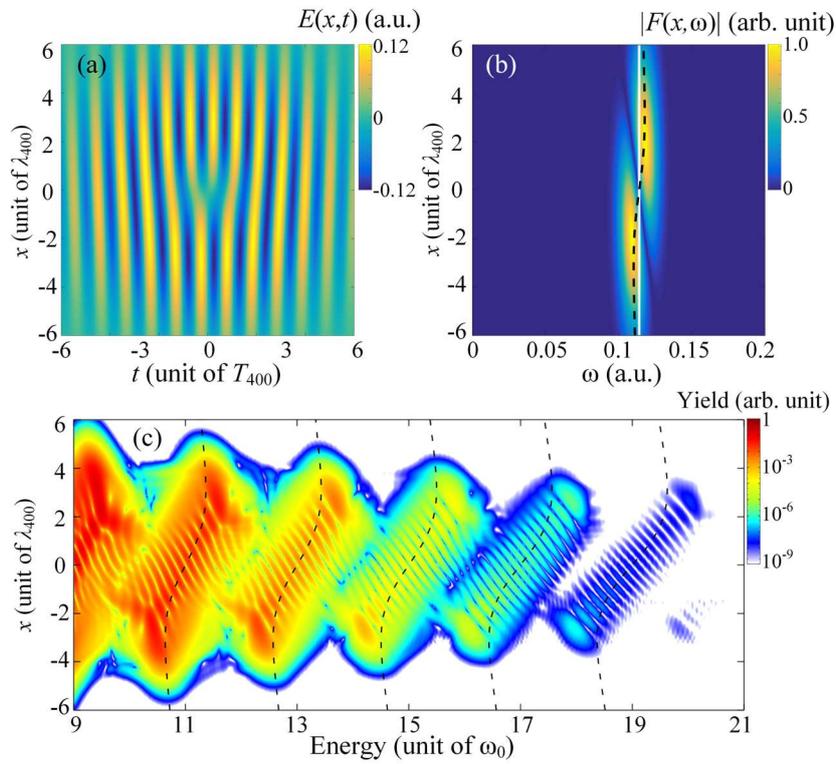

**FIG. 1.** (a) Spatiotemporal distribution of the electric field of the fundamental STOV. (b) The spatial chirp of the fundamental STOV. In (b), the pseudo-color map indicates the spatio-spectral weight of $|F(x,\omega)|$, the white line indicates the 400-nm angular frequency (i.e., $\omega_0 = 0.114$ a.u.), and the black dashed curve represents the local center angular frequency. (c), Spatial-resolved HHG spectrum driven by the fundamental STOV with hydrogen atoms.

**Figure 2**

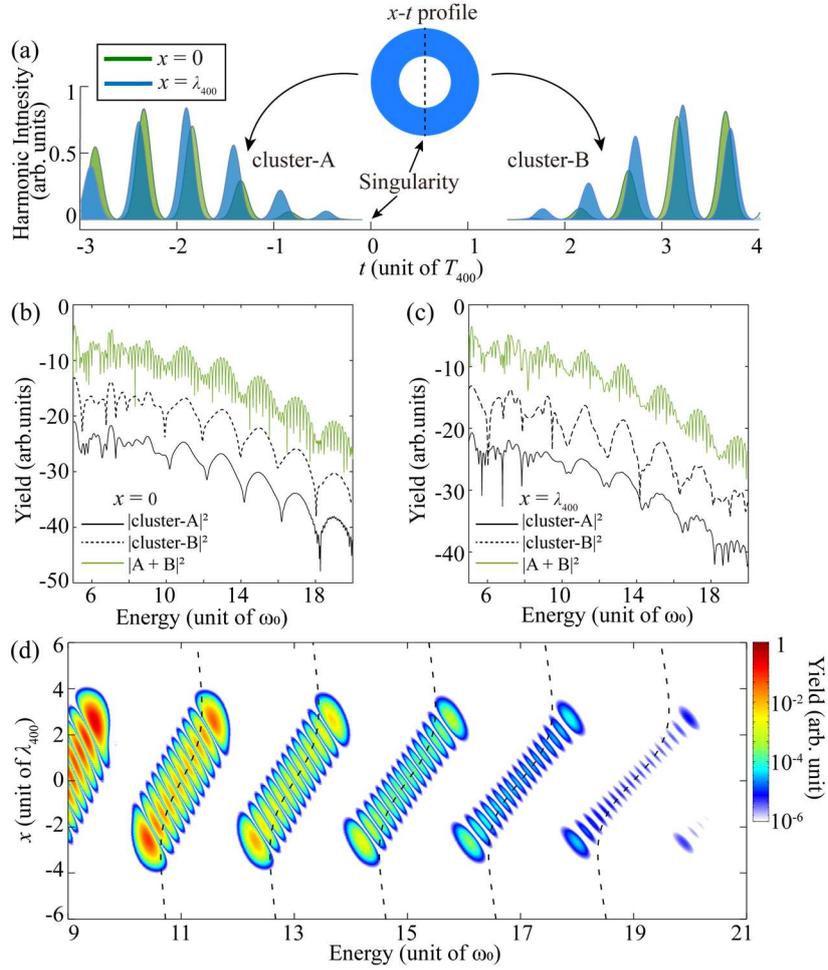

**FIG. 2.** (a) Illustration for the temporal character of HHG bursts. The harmonic bursts are separated into cluster-A and cluster-B in time domain by the phase singularity. (b) and (c) are HHG spectra from cluster-A, cluster-B, and their superposition, respectively. The $x$-positions for (b) and (c) are taken to be $x = 0$ and $x = \lambda_{400}$, respectively. (d) Spatial-resolved HHG spectrum from Eq. (5). In (d), the black dashed curves are the results from Eq. (4).

**Figure 3**

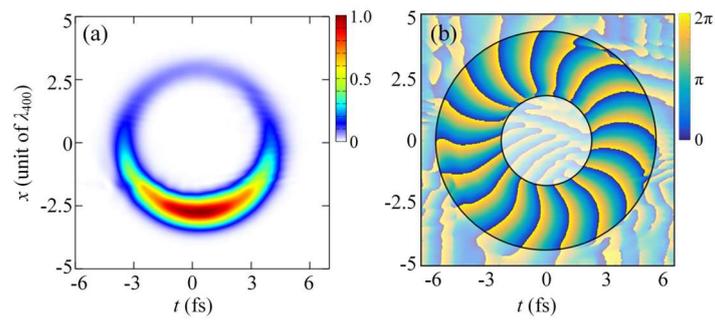

**FIG. 3.** The intensity (a) and phase (b) spatiotemporal profiles of the 17th harmonic.

# Figure 4

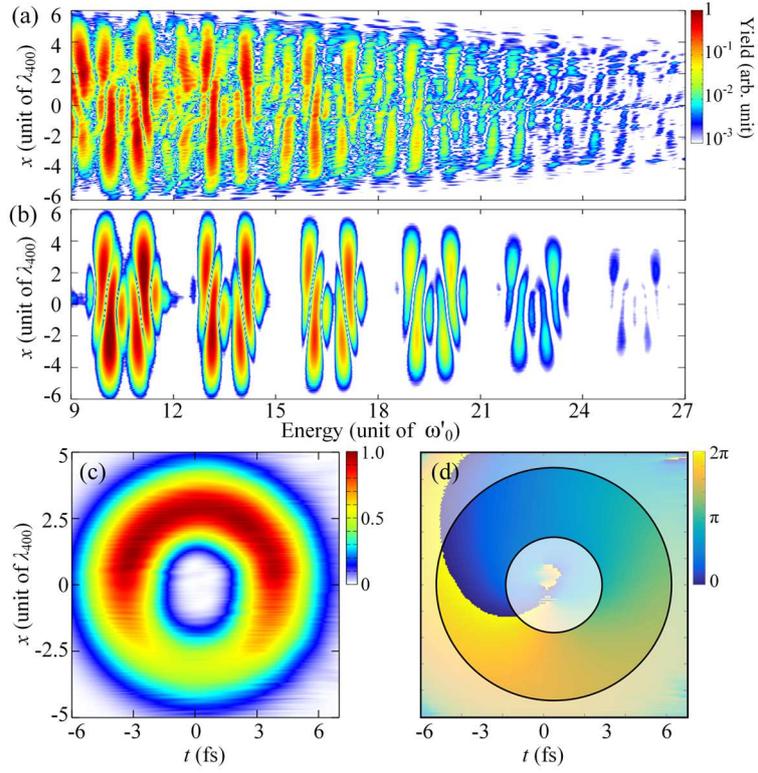

**FIG. 4.** (a) Spatial-resolved HHG spectrum driven by two-color counter-vorticity STOVs (400-nm + 800-nm). (b) Spatial-resolved HHG spectrum driven by two-color counter-spin and counter-vorticity STOVs. Here, $\omega'_0 = 0.057$ a.u. denotes the angular frequency of 800-nm field. The peak electric field strength of 400-nm and 800-nm light fields in (a) and (b) are both taken to be 0.12 a.u. (c) and (d) are the intensity and phase spatiotemporal profiles of the 17th harmonic, respectively.

**Figure 5**

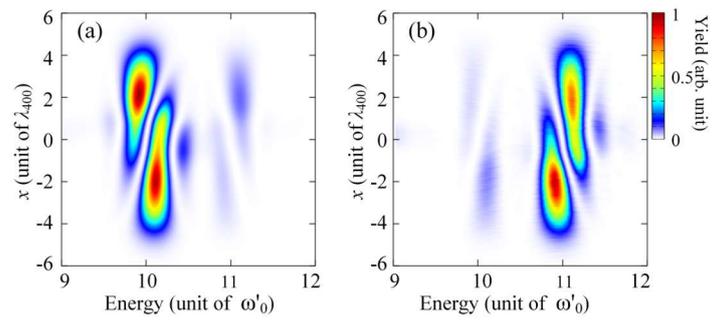

**FIG. 5.** Spatial-resolved HHG spectrum driven by two-color counter-spin and counter-vorticity STOVs at different intensity ratios ($I_{400}/I_{800}$). The intensity ratio is (a) 0.25 and (b) 4, respectively.